\documentclass[prb,twocolumn,showpacs,floatfix]{revtex4}
\usepackage{graphicx}
\usepackage{dcolumn}
\usepackage{bm}

\begin{document}
\title{Wave packet dynamics in a monolayer graphene}

\author{G.~M.~Maksimova, V.~Ya.~Demikhovskii, and E.~V.~Frolova}
\email{demi@phys.unn.ru} \affiliation{Nizhny Novgorod State
University,\\ Gagarin Ave., 23, Nizhny Novgorod 603950, Russian
Federation}

\date{today}

\begin{abstract}
The dynamics of charge particles described by Gaussian wave packet
in monolayer graphene is studied analytically and numerically. We
demonstrate that the shape of wave packet at arbitrary time
depends on correlation between the initial electron amplitudes
$\psi_1(\vec r,0)$ and $\psi_2(\vec r,0)$ on the sublattices $A$
and $B$ correspondingly (i.e. pseudospin polarization). For the
transverse pseudospin polarization the motion of the center of
wave packet occurs in the direction perpendicular to the average
momentum $ {\vec p_0}=\hbar \vec{ k_0}$. Moreover, in this case
the initial wave packet splits into two parts moving with opposite
velocities along  $ {\vec p_0}$. If the initial direction of
pseudospin coincides with average momentum the splitting is absent
and the center of wave packet is displaced at $t>0$ along the same
direction. The results of our calculations show that all types of
motion experience {\it zitterbewegung}. Besides, depending on
initial polarization the velocity of the packet center may have
the constant component $v_c=uf(a)$, where $u\approx 10^8~cm/s$ is
the Fermi velocity and $f(a)$ is a function of the parameter
$a=k_0d$ ($d$ is the initial width of wave packet). As a result,
the direction of the packet motion is determined not only by the
orientation of the average momentum, but mainly by the phase
difference between the up- and low- components of the wave
functions. Similar peculiarities of the dynamics of 2D electron
wave packet connected with initial spin polarization should take
place in the semiconductor quantum well under the influence of the
Rashba spin-orbit coupling.
\end{abstract}


\maketitle

\section{Introduction}

At last years the dynamics of wave packets in 2D electron gas and
other systems in solids including the phenomenon of {\it
zitterbewegung} (ZB) or trembling motion has been the subject of
numerous studies.
\cite{Shm,Schli,Sh8,Zaw1,Shen,Zaw2,Kat,Cser,Win,Trauz,Rus1,Rus2,Rus3}
Firstly the oscillatory motion analogous to the relativistic {\it
zitterbewegung} in two-dimensional systems with the structural and
bulk inversion asymmetry was investigated by Schliemann {\it et
al.}\cite{Schli} In the recent work by authors the detailed
studying of the electron wave packet dynamics in the semiconductor
quantum well under the influence of the Rashba spin-orbit coupling
was performed. \cite{Dem} It was shown (analytically and
numerically) that the initial wave packet splits into two parts
with different spin polarizations propagating with unequal group
velocities. It was demonstrated also that the splitting and
overlapping of wave packets leads to the damping of {\it
zitterbewegung}.

As well known, the electron {\it zitterbewegung} in relativistic
physics at first time was predicted by Schr\"odinger \cite{Schro}
(see also \cite{Bar}). This phenomenon is caused by the
interference between positive and negative energy states in the
wave packet. The frequency of ZB motion is determined by the gap
between these two states and when the amplitude of oscillations in a
particle position is of the order of the Compton wave length.
This phenomenon was discussed also in  Refs.[17-19].

In the papers by Rusin and Zawadzki \cite{Rus2,Rus3} the evolution
of the wave packet in a monolayer and bilayer graphene as well as
in carbon nanotubes was analyzed. The exact analytical expressions
for two components of wave function and average value of position
operator were found for bilayer graphene, which allowed to obtain
analytical results for the ZB of Gaussian wave packet. It was
shown that the transient character ZB in bilayer graphene is due
to the fact that wave subpackets related to positive and negative
electron energies move in opposite directions, so their overlap
diminishes with time. At the same time the dynamics of the wave
packets in a monolayer graphene in Ref.[12] was not investigated
fully.

In this work we present the detailed description of wave packet
evolution including the phenomenon of ZB of the packet center in a
monolayer graphene. We investigate the influence of the initial
pseudospin polarization on the dynamical characteristics of the
moving electron wave packet. Our analytically results are
illustrated by a graphic presentation.

The outline of this paper is as follows. The expression for the
wave function at $t>0$ is found in Sec.II for the arbitrary
initial pseudospin polarization. The time dynamics of Gaussian
wave packet and, in particular, the ZB phenomenon are described in
Sec.III for different correlations between the initial electron
amplitudes on the sublattices $A$ and $B$. We conclude with
remarks in Sec.IV.

\section{Basic equations}

Graphene is a single layer of carbon atom densely packed in a
honeycomb lattice. The two-dimensional  Hamiltonian describing its
band structure has the form \cite{Wall,Slon,Novos,Zhang}

$$\hat H=u\vec{\sigma}\hat{\vec{p}},\eqno(1)$$ where
$u\approx10^8$ $cm/s$, $\hat{\vec{p}}=(\hat p_x,\hat p_y)$ is the
momentum operator defined with respect to the centre of the valley
centered at the corner of the Brillouin zone with wave vector
$\vec K$. Pauli matrices $\sigma_i$ operate in the space of the
electron amplitude on two sites ($A$ and $B$) in the unit cell of
a hexagonal crystal. This internal degree of freedom plays a role
of a pseudospin. The Dirac-like Hamilton $\hat H$ determines the
linear dispersion relation

$$E_{p,s}=sup. \eqno (2)$$
Here $p=\sqrt{p_x^2+p_y^2}$, $s=1$ for the electron in the
conduction band and $s=-1$ for the valence band ("hole" branch of
quasiparticles). The corresponding eigenfunctions are given by

$$\varphi_{\vec{p},s}(\vec{r},t)=\frac{1}{2\sqrt{2}\pi\hbar}\exp(i\frac{\vec{p}\vec{r}}{\hbar}-i\frac{E_{p,s}t}{\hbar})
\pmatrix{1\cr s{\rm e}^{i\varphi}},\eqno(3)$$ with ${\rm
e}^{i\varphi}=\frac{p_x+ip_y}{p}$.

The time-evolution of an arbitrary initial state $\psi(\vec{r},0)$
in Shr\"odinger representation can be found with the help of
Green's function $G_{\mu \nu}(\vec{r},\vec{r^\prime})$

$$\psi_\mu (\vec{r},t)=\int G_{\mu
\nu}(\vec{r},\vec{r^\prime},t)\psi_\nu(\vec{r^\prime},0)d\vec{r^\prime},
\eqno(4)$$ where $\mu,\nu=1,2$ are matrix indices, corresponding
to the upper and lower components of $\psi(\vec{r},t)$. These
components are related to the probability of finding electron at the
sites of the sublattices $A$ and $B$ correspondingly. The standard
expression for Green's function is

$$G_{\mu \nu}(\vec{r},\vec{r^\prime},t)=\sum\limits_{s=\pm 1}\int
d\vec{p}~ \varphi_{\vec{p},s;\mu}(\vec{r},t)
\varphi_{\vec{p},s;\nu}^\ast(\vec{r^\prime},0). \eqno(5)$$
Using Eq.(3) for $\varphi_{\vec{p},s;\mu}(\vec{r},t)$ we find

$$\displaylines{G_{11}(\vec{r},\vec{r^\prime},t)=G_{22}(\vec{r},\vec{r^\prime},t)=\frac{1}{(2\pi
\hbar)^2}\times\cr\hfill\times\int
\exp(i\frac{\vec{p}(\vec{r}-\vec{r^\prime})}{\hbar})\cos(\frac{upt}{\hbar})d\vec{p},~~~\hfill\llap{(6)}\cr}$$

$$\displaylines{G_{21}(\vec{r},\vec{r^\prime},t)=G_{12}^\ast(\vec{r},\vec{r^\prime},t)=\frac{-i}{(2\pi
\hbar)^2}\times\cr\hfill\times\int \frac{p_x+ip_y}{p}
\exp(i\frac{\vec{p}(\vec{r}-\vec{r^\prime})}{\hbar})\sin(\frac{upt}{\hbar})d\vec{p},~~~\hfill\llap{(7)}\cr}$$

Let us represent the initial wave function by Gaussian wave packet
having the width $d$ and nonvanishing average momentum
$p_{0y}=\hbar k_0$

$$\psi(\vec{r},0)=\frac{f(\vec{r})}{\sqrt{|c_1|^2+|c_2|^2}}\pmatrix{c_1\cr c_2},\eqno(8a)$$

$$f(\vec{r})=\frac{1}{d\sqrt{\pi}}\exp(-\frac{r^2}{2d^2}+ik_0y),
\eqno(8b)$$ where coefficients $c_1$ and $c_2$ determine the
initial pseudospin polarization. We suppose that the packet width
$d$ is much greater than the lattice period and consequently $\psi
(\vec{r},0)$ is smooth enveloping function. We suppose also that
the most of the states in valence band are unfilled, that
corresponds to negative Fermi level located far from Dirac point
(see also \cite{RusZaw}). Substituting Eqs.(8a, 8b) in Eq.(4) and
using the expressions (6) and (7) we obtain

$$\psi_1(\vec{r},t)=\frac{1}{\sqrt{|c_1|^2+|c_2|^2}}(c_1\phi_1(\vec{r},t)-c_2\phi_2(-x,y,t)),
\eqno(9)$$

$$\psi_2(\vec{r},t)=\frac{1}{\sqrt{|c_1|^2+|c_2|^2}}(c_2\phi_1(\vec{r},t)+c_1\phi_2(\vec{r},t)), \eqno(10)$$
where, for notational convenience, $\phi_{1,2}(\vec{r},t)$ denote
the functions

$$\displaylines{\phi_1(\vec{r},t)=\int
G_{11}(\vec{r},\vec{r^\prime},t)f(\vec{r^\prime},0)d\vec{r^\prime}=
\frac{d{\rm
e}^{-(k_0d)^2/2}}{2\hbar^2\sqrt{\pi^3}}\times\cr\hfill\times\int
\exp(i\frac{\vec{p}\vec{r}}{\hbar}-\frac{p^2d^2}{2\hbar^2}+\frac{p_yk_0d^2}{\hbar})\cos(\frac{upt}{\hbar})d\vec{p},
~~~\hfill\llap{(11)}\cr}$$

$$\displaylines{\phi_2(\vec{r},t)=\int
G_{21}(\vec{r},\vec{r^\prime},t)f(\vec{r^\prime},0)d\vec{r^\prime}=
\frac{-id{\rm
e}^{-(k_0d)^2/2}}{2\hbar^2\sqrt{\pi^3}}\times\cr\hfill\times\int\frac{p_x+ip_y}{p}
\exp(i\frac{\vec{p}\vec{r}}{\hbar}-\frac{p^2d^2}{2\hbar^2}+\frac{p_yk_0d^2}{\hbar})\times\hfill\cr\hfill\times\sin(\frac{upt}{\hbar})d\vec{p}.
~~~\hfill\llap{(12)}\cr}$$

Using the cylindrical coordinates in Eqs.(11), (12) and
integrating over the angular variable, we have

$$\displaylines{\phi_1(\vec{r},t)=\frac{{\rm
e}^{-\frac{a^2}{2}}}{d\sqrt{\pi}}\times\cr\hfill\times\int\limits_{0}^{\infty}
{\rm e}^{-\frac{q^2}{2}}\cos(qt)J_0(q\sqrt{r^2-a^2-2iay})qdq,
~~~\hfill\llap{(13)}\cr}$$

$$\displaylines{\phi_2(\vec{r},\tau)=\frac{{\rm
e}^{-\frac{a^2}{2}}}{d\sqrt{\pi}}\frac{x+a+iy}{\sqrt{r^2-a^2-2iay}}
\times\cr\hfill\times\int\limits_{0}^{\infty} {\rm
e}^{-\frac{q^2}{2}}\sin(qt)J_1(q\sqrt{r^2-a^2-2iay})qdq,
~~~\hfill\llap{(14)}\cr}$$ where $J_0(z)$, $J_1(z)$ are Bessel
functions. For the sake of convenience we introduce in Eqs.(13),
(14) and everywhere below the dimensionless variables, measuring
the distance in the units of initial width of wave packet $d$ and
time in $d/u$ units. Besides, instead of the wave vector $k_0$ we
consider the parameter $a=k_0d$.

\section{{\it Zitterbewegung} of Gaussian wave packet with different pseudospin polarization}

Now we describe the time dynamics of Gaussian wave packets, in
particular, the ZB phenomenon and the influence of the initial
pseudospin polarization on the characteristics of trembling
motion.

i). Following Ref.[12] let us firstly consider the model
problem when the lower component of initial wave function is equal
to zero, i.e. the parameters $c_1=1$, $c_2=0$ in Eq.(8a). That
means that at the initial moment of time the electron probability is located at the sites of the sublattice $A$. It is not difficult to show that this
packet is formed by the states with positive and negative
energies. The relative weight of these states is equal to one. The
wave function for $t>0$ can be found using  Eqs.(9), (10):

$$\psi(\vec{r},t)=\pmatrix{\phi_1(\vec{r},t)\cr
\phi_2(\vec{r},t)},\eqno(15)$$ where the functions
$\phi_1(\vec{r},t)$, $\phi_2(\vec{r},t)$ are defined by
Eqs.(13),(14).

\begin{figure}
  \centering
  \includegraphics[width=80mm]{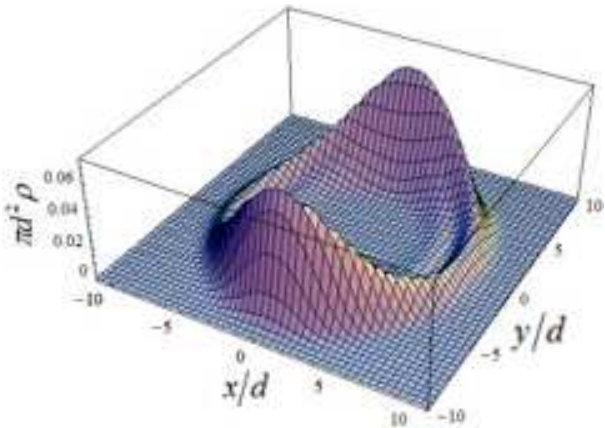}
\caption{(Color on line). The electron probability density
$\rho(\vec{r},t)=|\psi_1|^2+|\psi_2|^2$ for initial wave packet
determined by Eqs.(8a), (8b) with $c_1=1$ and $c_2=0$ for
$a=k_0d=1.2$ at the time $t=7$ (in the units of $d/u$).}
\label{fig1}
\end{figure}

In Fig.1 we represent the full electron density at the moment
$t=7$ for initial wave packet, Eq.(8b) with width $d=2$ $nm$ and
$k_0=0.6$ $nm^{-1}$. As one can see, at $t>0$ this packet splits
in two parts moving along $y$ axis with opposite velocities so
that the electron probability density is symmetrical with respect
to $y$: $\rho(x,y,t)=\rho(x,-y,t)$ (note that at the case $k_0=0$
the electron probability density has a cylindrical symmetry at all
time). For enough large time the width of both parts of the packet increases with time due to effect of dispersion. One can check that in this situation the contributions of two components of wave
functions $\psi_1(\vec r,t)$ and $\psi_2(\vec r,t)$ in full
electron density are equal. In other words the electron
probability distributes with the time on the sides of sublattice
$A$ and $B$.  Note at the same time $\rho(x,y,t)\neq \rho(-x,y,t)$
and the packet center oscillates along $x$ direction ({\it
zitterbewegung}).

\begin{figure}
  \centering
  \includegraphics[width=80mm]{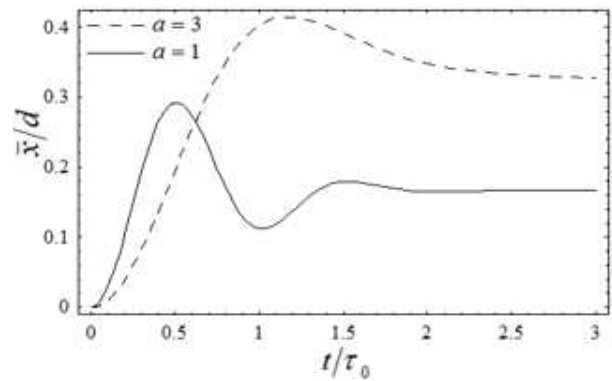}
\caption{The average coordinate $\bar x(t)$ versus time
($\tau_0=d/u$) for the wave packet with initial pseudospin
polarization along $z$ axis for two values of $a$.} \label{fig2}
\end{figure}

To analyze this motion we find the average value of position
operator. To do it, we use the momentum representation. The upper
$(C_1(\vec{p},t))$ and lower $(C_2(\vec{p},t))$ components of wave
function (15) in this representation can be easily obtained from
Eqs.(11), (12). After that the usual definition

$$\bar{\vec{r}}(t)=\sum\limits_{j=1}^{2}\int d\vec{p}~
C_j^\ast(\vec{p},t)i\hbar\frac{dC_j(\vec{p},t)}{d\vec{p}},
\eqno(16)$$

readily yields

$$\bar{y}(t)=0, \eqno (17a)$$

$$\displaylines{\bar{x}(t)=\frac{1-{\rm e}^{-a^2}}{2a}-{\rm
e}^{-a^2}\int\limits_{0}^{\infty}{\rm
e}^{-q^2}\cos(2qt)I_1(2aq)dq, ~~~\hfill\llap{(17b)}\cr}$$ where
$I_1(z)$ is a modified Bessel function of the first order. In
Eq.(17b) the integral term represents the {\it zitterbewegung}.
Note that average value $\bar x(t)$ depends on only one parameter
$a$ (in the dimensionless variables). The obtained functions $\bar
x(t)$ which describes the typical transient {\it zitterbewegung}
are plotted in Fig.2. After the oscillation disappears the center of
the packet is displaced by amount which equals to the first term
Eq.(15). In the case when the wave packet width is large enough
and the inequality $a=dk_0\gg 1$ takes place, Eq.(17b) reduces to
\cite{...}

$$\bar{x}(t)=\frac{1-{\rm e}^{-t^2}\cos(2at)}{2a}. \eqno (18)$$

As it follows from Eqs.(17), (18) for given initial polarization
of wave packet the ZB occurs in the direction perpendicular to the
initial momentum $p_{0y}=\hbar k_0$, just as for bilayer graphene
\cite{Rus2} and for the semiconductor quantum well in the presence
of the Rashba spin-orbit coupling \cite{Schli,Dem}. One can see
from Eq.(18) that the trembling motion has a transient character
as it was described in Refs.[12, 14] and at $t\gg 1$
$x(t)\rightarrow 1/2a$. We should notice that Eqs.(17b), (18)
coincide with corresponding formulas of Ref.[14]. This is because
the Hamiltonian for the system with Rashba-coupling

$$H_R=\frac{\hat{\vec{p}}^2}{2m}+\alpha(\hat{p}_y\hat\sigma_x-\hat{p}_x\hat\sigma_y),
\eqno(19)$$ where $\alpha$ is a Rashba coupling constant,
transforms into Hamiltonian for monolayer graphene, Eq.(1), if we
make the replacement in Eq.(19)

$$x\rightarrow -y^\prime,~~~y\rightarrow x^\prime,~~~ \alpha\rightarrow u,~~~ m\rightarrow
\infty. \eqno (20)$$

ii). Let us consider now the case when $c_1=c_2=1$, that is
pseudospin is directed along $x$ axis at $t=0$. Then from Eqs.(9),
(10)

$$\psi(\vec{r},t)=\frac{1}{\sqrt{2}}\pmatrix{\phi_1(\vec{r},t)-\phi_2(-x,y,t)
\cr \phi_1(\vec{r},t)+\phi_2(\vec{r},t)}.\eqno(21)$$

\begin{figure}
  \centering
  \includegraphics[width=80mm]{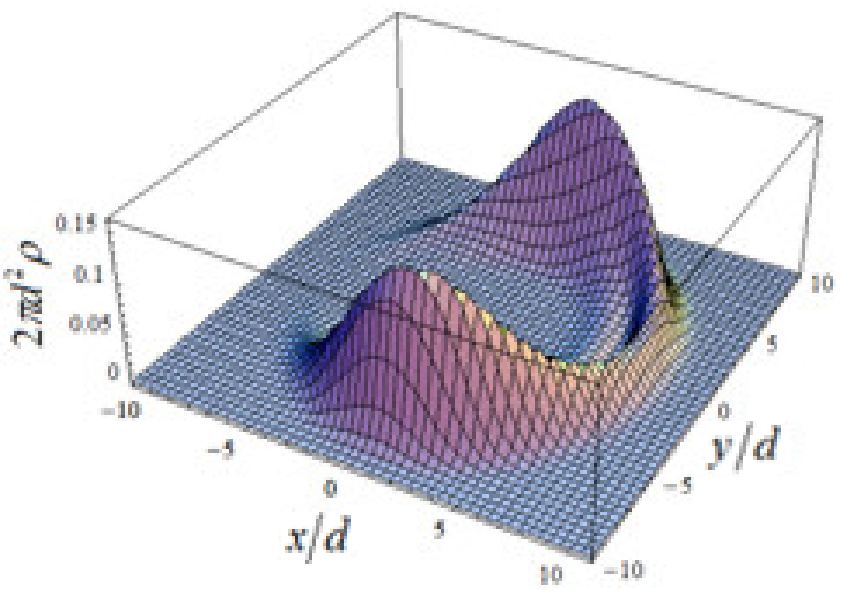}
\caption{(Color on line). The electron probability density
$\rho(\vec{r},t)=|\psi_1|^2+|\psi_2|^2$ for initial wave packet,
Eqs.(8a), (8b) with $c_1=1$ and $c_2=1$ for $a=k_0d=1.2$ at the
time $t=7$ (in the units of $d/u$).} \label{fig3}
\end{figure}

Fig.3 illustrates the corespondent electron probability density at
the time moment $t=7$ for initial wave packet, Eq.(8b), for the
same parameters as in Fig.1. One can see that the initial wave packet
at $t>0$, as in previous case, splits into two parts propagating
along $y$ in opposite directions so that the symmetry concerning
this axis, i.e. $\rho (x,y,t)=\rho (x,-y,t)$, retain during the
time (as the case i)). The distribution of the probability density
along $x$ axis clearly demonstrates that its maximum is displaced in the
positive direction that corresponds to the motion of the packet
centre along $x$ axis. The velocity of such motion $\bar
v_x=\frac{d\bar x}{dt}$ consists of both constant as well as
oscillatory parts. Really, a straightforward calculation yields
the average value of position operator $x$

$$\displaylines{\bar{x}(t)=\frac{1-{\rm
e}^{-a^2}}{2a^2}~t+\frac{{\rm
e}^{-a^2}}{2a}\times\cr\hfill\times\int\limits_0^\infty {\rm
e}^{-q^2} \sin(2qt)\frac{d}{dq}I_1(2aq)dq,
~~~\hfill\llap{(22)}\cr}$$ and $\bar{y}(t)=0$ like for the case
i). In Fig.4 we demonstrate the dependence $\bar x(t)$ for various
values of parameter $a$. When the parameter $a$ increases, the
amplitude of ZB and the period of oscillations decrease. At $a\gg
1$ we have from Eq.(22)

$$\bar{x}(t)=\frac{t}{2a^2}+\frac{1}{2a}{\rm e}^{-t^2}\sin(2at).
\eqno(23)$$

\begin{figure}
  \centering
  \includegraphics[width=80mm]{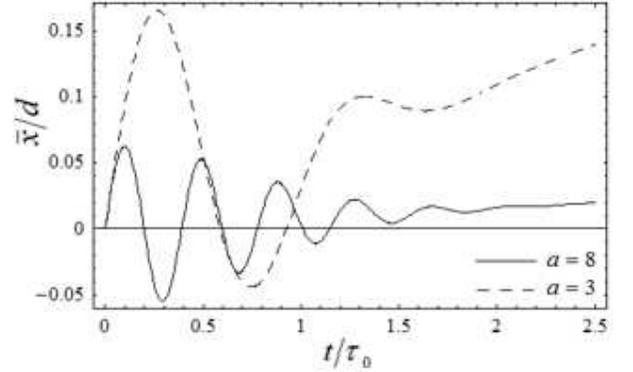}
\caption{The average coordinate $\bar x(t)$ versus time
($\tau_0=d/u$) for the wave packet with initial pseudospin
polarization along $x$ axis for different values of $a$.}
\label{fig4}
\end{figure}

We see that the character of motion of wave packet is changed. Now
the center of wave packet moves along $x$ direction with constant
velocity, which is determined by the first term in Eqs.(22), (23)
and performs the damping oscillations. This constant velocity has
a maximum value which is equal to $1/2$ (in the units of $u$) for
the narrow - width wave packet, or for the case $k_0=0$. When the
width of packet is increased the velocity of motion of its centre
is decreased. The frequency and amplitude of the {\it
zitterbewegung} for $a\gg 1$ are the same as in the case i).
However, the first term in Eq.(22) corresponding to the motion of
wave packet with constant velocity reduces the effect of ZB at
least for $a\lesssim 1$ (Fig.4).

It is not difficult to show that as in the other two-band systems
the phenomenon of ZB in graphene is a result of an interference of
states corresponding to positive and negative eigenenergies of
Hamiltonian, Eq.(1). For wide enough packet $a=k_0d\gg 1$ and at
time $t>1$ when the ZB disappears two split parts of initial
wave packet (see Fig. 3) move along $y$ axis with opposite
velocities $u$ and $-u$. In this situation the subpackets moving in
the positive and negative directions consist of the states with
positive and negative energies correspondingly.

iii). When the initial pseudospin is along $y$ axis the wave
function at $t>0$ has the form

$$\psi(\vec{r},t)=\frac{1}{\sqrt{2}}\pmatrix{\phi_1(\vec{r},t)-i\phi_2(-x,y,t)\cr
i\phi_1(\vec{r},t)+\phi_2(\vec{r},t)}.\eqno(24)$$

\begin{figure}
  \centering
  \includegraphics[width=85mm]{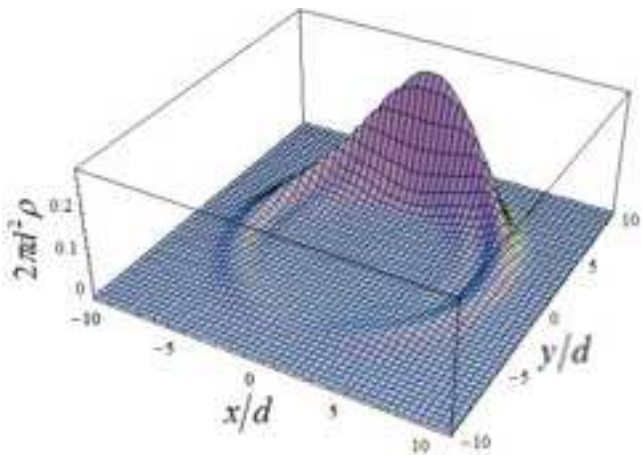}
\caption{(Color online). The electron probability density
$\rho(\vec{r},t)=|\psi_1|^2+|\psi_2|^2$ for initial wave packet,
Eqs.(8a), (8b) with $c_1=1$ and $c_2=i$ for $a=k_0d=1.2$ at time $t=7$ (in the units of $d/u$).} \label{fig5}
\end{figure}

In Fig.5 the full electron density for the same moment of time and for the same parameters as in previous cases is shown. As one
can see, the initial wave packet does not split into two parts at
$t>0$ unlike in the cases i) and ii). This result is confirmed by the
straightforward calculations. Indeed, one can show that the
eigenenergy states corresponding to propagation in the positive
direction along $y$ axis give the dominant contribution in total
wave function, Eq.(24). For wide packets ($a\gg 1$) almost all of
these states belong to the positive branch of energy.

\begin{figure}
  \centering
  \includegraphics[width=80mm]{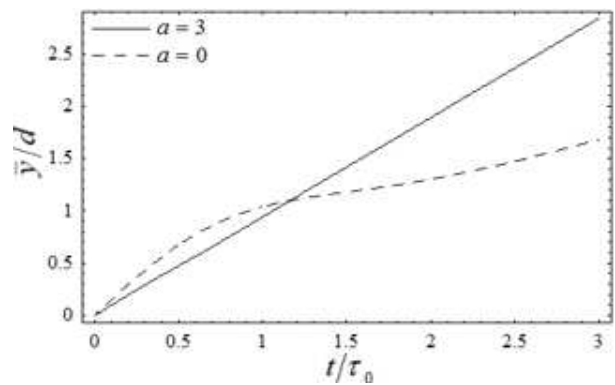}
\caption{The average coordinate $\bar y(t)$ versus time
($\tau_0=d/u$) for the wave packet with initial pseudospin
polarization along $y$ axis for different values of $a$.}
\label{fig6}
\end{figure}

The results of calculations of average values of $x$ and $y$ for
this polarization lead to

$$\bar{x}(t)=0, \eqno(25)$$

$$\displaylines{\bar{y}(t)=(1-\frac{1}{2a^2}+\frac{{\rm
e}^{-a^2}}{2a^2})~t+\frac{{\rm
e}^{-a^2}}{2a}\times\cr\hfill\times\int\limits_0^\infty {\rm
e}^{-q^2} \sin(2qt)I_1(2aq)\frac{dq}{q}.
~~~\hfill\llap{(26)}\cr}$$ Thus in the considering case the wave
packet propagates along $y$ axis and the {\it zitterbewegung} has
the "longitudinal" character \cite{formula}.

Note that in a numerical work  \cite{Thal} by Thaller the authors have observed
similar oscillatory behavior of the expectation value of the
position operator for one - dimensional relativistic electron in
vacuum. In Ref.[18] it was also shown that apart from the rapid
oscillation, the wave packet drifts slowly even when its average
momentum is zero.

In Fig.6 we represent the dependence $\bar y(t)$ for different
values of parameter $a$. As one can see, even at zero value
of $a$ the oscillations are absent. In this case, as it follows
from Eq.(26), the drift velocity is equal to $1/2$ (in the units of
$u$). As above, Eq.(26), takes more simple form at $a\gg 1$

$$\bar{y}(t)=t+\frac{1}{4a^3}{\rm e}^{-t^2}\sin(2at). \eqno(27)$$

Comparing Eqs.(18), (23), (27), we see that the amplitude for the
"longitudinal" {\it zitterbewegung} is much smaller than the amplitude of
"transverse" {\it zitterbewegung}. This fact can bee seen as a
consequence of special form of the initial wave function, which in
the given case consists of (at $a\gg 1$) the states with positive
energy mostly. That makes the interference between the positive and
negative components difficult, i.e. decreases the ZB. Moreover, at
any values of the parameter $a$ the integral term in Eq.(26),
corresponding to the oscillating motion, is negligible in
comparison with the first term, and one may neglect the effect of
the "longitudinal" ZB.

As was demonstrated above, the direction of the average velocity
depends not only on module of the components $\psi_1(\vec{r},0)$
and $\psi_2(\vec{r},0)$, but also on their phases. In general case
for the initial Gaussian packet

$$\psi(\vec{r},0)=\frac{f(\vec{r})}{\sqrt{2}}\pmatrix{1\cr {\rm e}^{i\varphi}},\eqno(28)$$
the probability density becomes asymmetric and the average
position operator has two components

$$\bar {\bf r}(t)=\bar x(t)\cos\varphi ~\vec{{\rm e}}_x+ \bar y(t)\sin\varphi~ \vec{{\rm
e}}_y ,\eqno(29)$$ where $\varphi$ is an arbitrary phase
difference between the up and low components of wave function and
$\bar x(t)$, $\bar y(t)$ are determined by Eqs.(22), (26). For
illustration we show in Fig.7 the electron probability
density obtained for the initial packet, Eq.(28), with
$\varphi=\pi/4$.

\begin{figure}
  \centering
  \includegraphics[width=80mm]{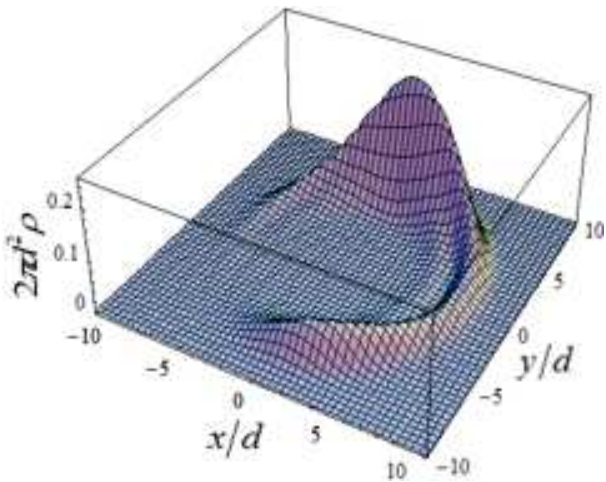}
\caption{(Color online). The electron probability density
$\rho(\vec{r},t)=|\psi_1|^2+|\psi_2|^2$ for initial wave packet,
Eq.(8a), (8b) with $c_1=1$ and $c_2={\rm e}^{i\pi/4}$ for
$a=k_0d=1.2$ at the time $t=7$ (in the units of $d/u$).}
\label{fig5}
\end{figure}

\begin{figure}
  \centering
  \includegraphics[width=80mm]{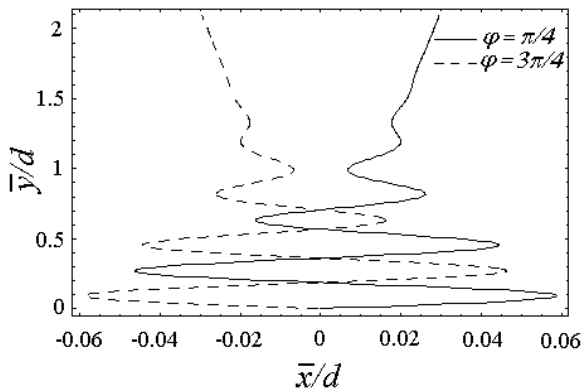}
\caption{The trajectories of the center of electron wave packet
described by Eq.(28) for two initial phases $\varphi=\pi/4$ and
$3\pi/4$. The parameter $a=6$.} \label{fig8}
\end{figure}

It is clear that the phase $\varphi$ determines the direction of
the average velocity of the packet center. Using the expression
for velocity operator $\hat{\vec v}=u\vec\sigma$ and Eq.(28) we
obtain (in the dimensionless variables) at $t=0$:
$$
\overline{v}_x(0)=\cos\varphi,\quad
\overline{v}_y(0)=\sin\varphi.\eqno(30)
$$
The components of the velocity for a large enough time, when the
trembling motion stops, can be found from Eqs.(22),(26) and (29)
for arbitrary parameter $a$
$$
\overline{v}_x=\frac{1-\exp(-a^2)}{2a^2}\cos\varphi,\;
\overline{v}_y=\bigg(1-\frac{1}{2a^2}+\frac{\exp(-a^2)}{2a^2}\bigg)\sin\varphi.\eqno(31)
$$
In particular, as it follows from Eq.(31) for $a\ll 1$, the
direction of the motion of wave packet center at large time
coincides with the initial one, Eq(30). In other limiting case
$a\gg 1$ (and for not too small $\varphi$) asymptotic direction of
the average velocity is along $0Y$ axis, i.e. along the average
momentum of wave packet $p_y=\hbar k_0$. Thus, by changing the
initial phase $\varphi$, one can govern the packet motion and
consequently the direction of dc current. To illustrate this, we
plot in Fig.8 the packet center trajectories for two initial
phases: $\varphi=\pi/4$ and $3\pi/4$. Note that the packet motion
with the constant velocity predicted above (see Eqs.(22), (26))
should lead to the existence of the direct current in the
corresponding direction.

To check our formalism let us consider the plane wave as the starting point. In this case it is easy to obtain the expression for
the average value of electron velocity $\bar{\vec{v}}(t)$. Really, in
the Heisenberg picture the kinetic velocity is (in the dimensional
variables)

$$\hat{\vec{v}}(t)=\frac{1}{i\hbar}[\vec{r},\hat{H}]=u\vec{\sigma}(t),
\eqno(32)$$ where

$$\frac{d\vec{\sigma}}{dt}=\frac{1}{i\hbar}[\vec{\sigma},\hat{H}]=\frac{2u}{\hbar}[\hat{\vec{p}}\times\vec{\sigma}].
\eqno(33)$$

In these equations $\vec{p}(t)=\vec{p}(0)$. Let the initial
momentum $p_{oy}=\hbar k_0$. Then, using the solutions of Eqs.(32),
(33) we find

$$\bar v_x(t)=u\bar \sigma_z(0)\sin\omega t+u\bar \sigma_x
(0)\cos\omega t, \eqno(34a)$$

$$\bar v_y (t)=u\bar \sigma_y (0), \eqno(34b)$$ where
$\omega=2uk_0$ and $\sigma_i(0)=\sigma_i$ - Pauli matrixes
($i=1,2,3$). So, if in the initial state pseudospin is along $z$
direction, i.e. $\bar \sigma_z(0)=1$ (case i)) we obtain from
Eq.(34a) that $\bar v_x (t)=u\sin\omega t$ which leads to

$$\bar{x}(t)=const-\frac{u}{\omega}\cos \omega t.
\eqno(35)$$ Returning to the original variables in Eq.(18) and
setting $d=\infty$ we see that this expression coincides with
Eq.(35). We get similar results also for other initial polarizations.

\section{Concluding remarks}

We have studied the quantum dynamics of charge particles
represented by Gaussian wave packets in two-dimensional single
layer of carbon atoms (graphene). We investigated numerically also
the spatial evolution of the initial wave packet and demonstrated
the effect of the packet splitting for the pseudospin polarization
perpendicular to the average momentum. The analytical expressions
for the average values of position operators were obtained. These
expressions clearly demonstrate that the evolution of wave function is
accompanied by the {\it zitterbewegung} and strongly depends on
the initial pseudospin polarization.  In particular, if the
initial pseudospin polarization coincides with initial average
momentum, the packet center moves and exhibits the ZB along the
same direction. In this case the second term in Eq.(26) describing
the longitudinal oscillations (the "longitudinal" ZB) is
essentially smaller than the first one connected with the motion
with constant velocity. As for other systems with two-band
structure \cite{Dem,Rus2,Thal}, the ZB in monolayer graphene has
a transient character.

It was also predicted that apart from the packet center exhibits
the trembling motion it can move with constant velocity (for the
pseudospin polarization along $x$ and $y$ axis). The direction of
this velocity depends on not only the orientation of average
momentum $\vec p_0$, but also on the module of the components
$\psi_1(\vec{r},0)$, $\psi_2(\vec{r},0)$ and the differences of
their phases (see Eqs.(28),(31)).

All above calculations have been done for the $\vec{K}$ point of
the Brillouin zone in graphene. Similar results can be found for
initial wave packet with wave vector $\vec{k}$ in the valley
centered in inequivalent point $\vec{K'}$. The Dirac Hamiltonian
around $\vec{K'}$ point can be written as\cite {text1}

$$H_{K'}=u\pmatrix{0 & -\hat{p}_x-i\hat{p}_y\cr
-\hat{p}_x+i\hat{p}_y & 0},\eqno(36)$$

This expression can be obtained from Hamiltronian around $\vec{K}$
point given by Eq.(1) by replacement
$\hat{p}_x\rightarrow-\hat{p}_x$. Thus values $\bar{x}(t)$ for the
wave packet of different polarizations (and corresponding
components of velocity) change sign while $\bar{y}(t)$ remain
unchanged (see also\cite{RusZaw}).

In conclusion we would like to stress that the packet motion with
the constant velocity (see Eqs.(22), (26)) leads to the appearance
of the dc current. For the experimental detection of this current
one needs sensitive current meters. Experimental observation of
trembling motion is currently a more difficult task since it is
necessary to use femtosecond techniques. \cite{Zaw2,Rus2} If new
methods of formation of wave packets with different pseudospin
polarizations will be elaborated then their trajectories and
spatial separations can be observed probably with the help of
devices with quantum point contacts and gates (see for example
\cite{Cas}).

\section*{Acknowledgments}
This work was supported by the program of the Russian Ministry of
Education and Science "Development of scientific potential of high
education" (project 2.1.1.2363).


\begin{thebibliography}{28}
\bibitem{...} The details of the calculation of asymptotic formula
similar to Eq. (18) are represented in Ref. [14].
\bibitem{formula} It is interesting to note that 1D wave packet
$\psi(y,0)=F(y)\pmatrix{1\cr i},$ where $F(y)$ is arbitrary
function, propagates along $y$ direction without changing of its
form $\psi(y,t)=F(y-ut)\pmatrix{1\cr i}.$
\end{thebibliography}
\end{document}